
\hoffset = .75   in
\voffset = .75   in
\centerline{\bf Van Hove Exciton-Cageons and High-T$_c$ Superconductivity:
VIIID}
\centerline{\bf Solitons and Nonlinear Dynamics}
\smallskip
\centerline{R.S. Markiewicz}
\centerline{Physics Department and Barnett Institute, Northeastern University,
Boston, MA 02115}
\bigskip
\par\noindent
Running Title: {\bf Solitons and Nonlinear Dynamics}
\bigskip
\par
The low-temperature orthorhombic (LTO) phase transition in
La$_{2-x}$Sr$_x$CuO$_4$ can be interpreted as a dynamic Jahn-Teller
effect, in which the degenerate electronic states are associated with the
large densities of states at the two van Hove singularities.  The equations
describing this phase are strongly nonlinear.  This paper illustrates some
consequences of the nonlinearity, by presenting a rich variety of exact
nonlinear wave solutions for the model.
\par
Of particular interest are soliton lattice solutions: arrays of domain walls
separating regions of local low-temperature tetragonal (LTT) symmetry.  These
arrays have a {\it macroscopic} average symmetry higher than LTT.  These
lattices can display either orthorhombic (`orthons') or tetragonal
(`tetrons') symmetry, and can serve as models for a microscopic description
of the dynamic JT LTO and high-temperature tetragonal phases, respectively.
\bigskip
\noindent{\bf 1. Introduction}
\bigskip\par
The two-dimensional van Hove singularity (vHs) was initially introduced as a
model[1] to study competition between superconductivity and structural
instability.  Its possible relevance to high-T$_c$ superconductivity was
pointed
out by Hirsch and Scalapino[2], who showed that, because of the logarithmically
diverging density of states (dos), the form of the BCS equation is modified in
a vHs superconductor, leading to much higher values of $T_c$.  When the
high-T$_c$ superconductors La$_{2-x}$Ba$_x$CuO$_4$ (LBCO) and La$_{2-x}$Sr$_
x$CuO$_4$ (LSCO) were first discovered, it was noted that their Fermi surface
falls close to a vHs, and it was proposed that the vHs might be responsible for
both the high $T_c$ values, as well as the structural transition from the
high-temperature tetragonal (HTT) phase to the low-temperature orthorhombic
(LTO) phase[3].
\par
However, the LTO phase was found not to split the degeneracy of the two vHs[4],
and moreover did not have a large effect on the electronic properties of LSCO,
so it was concluded that the LTO transition had nothing to do with the vHs.
Now the absence of a structural transition does not rule out a role for the
vHs in promoting superconductivity.  Indeed, the $T_c$ enhancement depends
mainly on the dos peak -- i.e., on the $q=0$ charge susceptibility, and not the
inter-vHs $q=2k_F$ susceptibility[5].  However, Phillips has shown[6] that
a simple vHs model cannot explain many details of, e.g., the doping
dependence of $T_c$.  Most of these problems could be overcome if the vHs also
produced short-range structural disorder[7], but, according to Phillips,
``there
is no experimental evidence which would indicate any relation between the major
lattice instabilities'' and the vHs[6].
\par
In fact, such evidence is starting to accumulate.  First, there is a second
structural phase transition in LBCO, this time to a low-temperature tetragonal
(LTT) phase[8], which does have a large impact on electronic properties and
essentially destroys superconductivity.  Moreover, the transition is optimized
at a fixed concentration of holes[9], and has the correct symmetry to split the
vHs degeneracy[10].  Secondly, a careful reexamination of the LTO-HTT phase
boundary has showed that superconductivity is clearly associated with the LTO
phase[11].  Finally, it was recently proposed[12-14] that the LTO phase
transition is indeed driven by the vHs, but the LTO phase is {\it dynamic}, and
the average macroscopic symmetry is not representative of the local
microscopic state.
\par
This new model describes the structural phase transitions in terms of the
interplay between the two electronic saddle points (vHs's) and the optical
phonons associated with tilting modes of the oxygen octahedra, with subsidiary
roles played by shear strains and acoustic phonons.  The LTT phase
involves an essentially static band Jahn-Teller (JT) effect[13], in which the
two vHs's play the role of the degenerate electronic states.  As the
temperature
is raised, there are large fluctuation effects, and the LTO and HTT phases can
be described as dynamic and disordered JT phases, respectively[14].
\par
The model Hamiltonian is an extension of the Labb\'e-Friedel model for the A15
compounds[15,16].  It contains significant nonlinearities; the purpose of the
present manuscript is to explicitly display some of these nonlinear features.
Analysis of these nonlinear waves should prove useful in understanding the
short-range order present in LBCO above the LTT transition, and as fluctuations
in other high-T$_c$ cuprates[17], and in understanding the role of
anharmonicity in enhancing T$_c$.  The present paper introduces a concrete
microscopic model which displays the properties postulated for the dynamic JT
phase.
\par
Many of the anomalous features are shared by a larger class of materials, the
ferroelastics[18].  In addition to the high-$T_c$ cuprates, this class includes
a number of related perovskite compounds and ferroelectrics, including SrTiO$_
3,$ and BaTiO$_3$, and the martensitic phases of the A15's.  Ferroelasticity is
in general suggestive of collective dynamic JT effects, with strong
electron-phonon coupling.  Hence, the present model can be applied to
transitions within this larger class of materials.  Indeed, the model allows
significant electron-phonon coupling even for a filled band, if the Fermi level
falls in a gap between two hybridized bands[13].
\par
The paper is organized as follows.  Section 2 shows that corner-sharing of
octahedra greatly restricts the possible low-lying states of the lattice, and
requires the introduction of defects (solitons) to expand the variety of
configurations.  A possible model for domain-wall solitons is introduced, and
it is shown how this state can provide a microscopic picture for the dynamic JT
state of LSCO.  In Section 3, the combined electron-phonon equations of motion
are introduced, and reduced to a form which is convenient for analyzing the
nonlinear wave solutions, both for waves propagating along 100 ($x$ or $y$
axes)
or 110 (at $45^o$) directions.  Section 4 illustrates the variety of possible
wave solutions, both dynamic and static.  The low energy waves include both
phason-like modes (weak, periodic modulation of the axis of tilt) and
soliton-like solutions.  These latter can be either isolated domain walls
between LTT domains, or arrays of solitons, which display an average
macroscopic symmetry which is greater than the true microscopic (LTT) symmetry.
The macroscopic symmetry can be either orthorhombic (the corresponding waves
are
called `orthons') or tetragonal (`tetrons').  In Section 5, these results are
applied to LSCO.  The material parameters are estimated, and the energies of
the resulting solitons are calculated.  Soliton array solutions are predicted
to
arise for 100-type waves, including static arrays.  The energy of individual
domain walls is calculated, and compared to microscopic calculations (Appendix
II).  The relation of the various
types of soliton arrays (tetron, orthon, isolated domain walls) and the
observed
phase transitions (HTT $\rightarrow$ LTO $\rightarrow$ LTT) in LSCO/LBCO is
discussed.  In Section 6, it is suggested that the solitons may be charged, and
hence related to domain wall phases postulated earlier in doped LSCO.  Finally,
Section 7 summarizes the conclusions of this work.
\bigskip
\noindent{\bf 2. Model of Solitons}
\smallskip\par
The essential feature of the dynamic JT effect is that the global symmetry
is not indicative of the local symmetry.  Such a situation is found in other
materials, in particular in the several phase transitions of BaTiO$_3$[19].
In the high temperature phases of BaTiO$_3$, the sample is composed of
microscopic domains of the low temperature (distorted) phase, but with
different polarization axes in each domain.  As the temperature is raised,
the axes point successively over two, four, or all eight possible orientations,
so that the sample as a whole appears to have higher and higher symmetry.
These domains have been directly observed via streaking in electron diffraction
patterns[20], and similar streaking has been observed in high-$T_c$
compounds[21].
\par
The tilting in these materials has often been described at the mean-field level
in terms of a pseudospin or Potts model, replacing the allowed tilts with
equivalent spin states.  A closely related mean-field treatment was applied to
LSCO in Ref. [14].  A shortcoming of this procedure is that it overlooks the
problems associated with corner or edge sharing of octahedra[22].  This corner
sharing leads to a long-range
interaction between tilts, and requires the introduction of macroscopic defects
(solitons as domain walls) to break up the resulting LTT-type order.
\par
Let us illustrate this effect in LSCO.  Consider a single octahedron,
tilted about the x-axis (as in the LTT ground state).  Since neighboring
octahedra are corner-sharing, its nearest neighbors along the y-axis must be
counter-tilted about the x-axis (again, as in the LTT phase).  By induction,
the whole row must be similarly tilted, in the absence of some form of
excitation -- either an anti-tilt along x or a crossover to a tilt about y.
Similarly, if the entire row is tilted about x, the next row along the x-axis
cannot be tilted about the y-axis, without tipping an oxygen which is shared
with the original row.  Hence, the second row would also be tilted about the
x-axis.  By induction, every row would be tilted about the x-axis.  Hence, in
the absence of defects, the ground state would have a quasi-LTT pattern.
To lowest order, there is only a weak interaction between the tilt direction
on adjacent rows, so the low energy physics would (in the absence of defects)
be that of rows of x-tilted octahedra, with two possible states for each row
(e.g., any particular octahedron could have either a + or -- tilt about x,
which
would fix the tilts of all other octahedra on that row).  Hence, this physics
would be in the universality class of the one-dimensional Ising model.
\par
Such one-dimensional chains have very low entropy: for an $N\times N$ array of
atoms there are $N^2$ atoms, but only $N$ chains, each of which has only two
possible orientations, for a total of $2^N$ configurations.  To increase the
entropy significantly, it is necessary to introduce macroscopic defects into
the system.  For instance, it would be preferable to
have domains in half of which the tilts lie along the y-axis, in the other
half along the x-axis.  These domains of LTT-like phase must be separated by
domain walls (solitons).  The purpose of the present manuscript is to use
the nonlinear equations of the system to develop a model for these domain
walls.
If the wells are deep, the domain walls
are likely to be narrow -- perhaps a single cell in width.  If the wells are
shallow (or at high temperatures) the walls can be broad as the octahedral tilt
changes gradually.  The average cell size will be given by the balance between
surface tension and entropy.
\par
Figure 1 suggests a possible (narrow wall) soliton model of the LTO phase.
The circles in Fig. 1 represent the apical oxygens above the CuO$_2$ plane,
which would be centered immediately above the Cu in the undistorted HTT
phase.  The arrows illustrate the instantaneous local distortion of the
apical oxygen's position in the tilted state.  The figure illustrates two
LTT-like domains, separated by a dynamic domain wall, with rotating tilt
direction, for motion either along a (110) direction (Fig. 1a) or along a
(100) direction (Fig. 1b).  In the (110) wave, the domains are composed of
diagonal stripes, alternating in tilt along the x- and y-axes.  The line of
atoms along the domain boundary cannot, because of corner sharing, be in a pure
LTT tilt, but will assume a (dynamic) average tilt, as illustrated by the
arcs of circles.  If the octahedra within each cell have pure LTT tilts, the
structure in the figure would not directly produce any LTO strain, $\epsilon_{
xy}$.  However, there is a pronounced asymmetry to the tilts along directions
parallel and perpendicular to the domain boundaries.  Within each row parallel
to the boundary, the tilt is uniquely either up or down, so that the entire
domain can readily assume the corrugated pattern characteristic of the LTO
phase (i.e., strong coupling to a $\epsilon_{xy}$ strain.  On the other hand,
in the perpenducular direction, the up and down tilted rows are interchanged
each time a soliton wall is crossed, making it impossible to generate a
global corrigation in this direction.  Moreover, the octahedra comprising
the soliton {\it already} lead to a net shear strain in the easy direction.
\par
Thus, the proposed configuration (1) has domains of local LTT symmetry,
(2) arranged in such a pattern as to readily couple to a corrugating shear
along a preferred easy axis, to give a {\it macroscopic} LTO symmetry.
Finally, (3) the domain walls already provide the appropriate shear strain,
so that $\epsilon_{xy}$ should be approximately proportional to the
density of solitons.
\par
It is very often convenient, in dealing with structural phase transitions, to
think in terms of pseudospins, since the magnetic analogs are generally much
better developed.  Within such a pseudospin model, the solitons have a very
natural interpretation: the two kinds of domain correspond to $S_z>0$ and $S_z
<0$, while the soliton is simply a N\'eel wall, with $S_x\ne 0$ inside the
wall.
\bigskip
\noindent{\bf 3. Real-Space Hamiltonian: Intracell and Intercell Coupling}
\smallskip\par
\bigskip
\noindent{\bf 3a. Tilt Equation of Motion}
\smallskip\par
In the present paper, the Hamiltonian will be a simplified form of that
introduced in Ref. [14] (Eq. 25 of that paper).  The coupling to strains and
umklapp electron sacttering were found to be relatively unimportant, and hence
will be neglected.  Furthermore, it will be assumed that the octahedra are
all tilted, but in random directions, so the pseudorotation operators have
the form $R_1=\bar Rcos\phi$, $R_2=\bar Rsin\phi$, with $\bar R$ constant.
The Hamiltonian then becomes
$$H=\sum_l\Bigl({1\over 2}M\bar R^2\dot\phi^2(\vec l)+H_{epl}$$
$$+{\Gamma_0\over 4}\sum_{\eta}[({R_1({\vec l})-R_1({\vec l}+
\eta a\hat y)\over 2})^4+({R_2({\vec l})-R_2({\vec l}+\eta a\hat x)\over 2})^4]
+{\Gamma_0^a\over 2}[R_1^4({\vec l})+R_2^4({\vec l})]
+{\Gamma_2\over 2}R_1^2({\vec l})R_2^2({\vec l})\Bigr),\eqno(1a)$$
$$H_{epl}=-\tilde\alpha_-^{e}|
\sum_{\eta}[({R_1({\vec l})-R_1({\vec l}+\eta
a\hat y)\over 2})^2-({R_2({\vec l})-R_2({\vec l}+\eta a\hat x)\over 2})^2]|,
\eqno(1b)$$
where $\eta$ is summed over $\pm 1$.  It should be noted that $R_i$ does {\it
not} have the dimensions of a length[13]; rather
$$R_i^2={3\over 2}m_OZ^2,$$
where $m_O$ is the mass of an oxygen atom and $Z$ is the displacement of the
planar oxygen above/below the CuO$_2$ plane caused by the pseudorotation $R_i$.
This means that the parameter $M$ is dimensionless,
$$M={4\over 3}(1+\eta_{ap}^2)\simeq 3.48,$$
where $\eta_{ap}\simeq 1.27$ is the ratio of the Cu-apical O distance to the
Cu-planar O distance.
\par
In Eq. 1b, $H_{epl}$ is the electron-phonon coupling term in the $l$th cell,
and
the absolute value arises from the JT effect[14].
This absolute value sign leads to unnecessary complications in the
analysis, since the model is only approximate to begin with, so a further
approximation is introduced.  This approximation essentially amounts to
Fourier analyzing the electron-phonon term in $cos(4n\phi )$, and retaining
only the lowest order terms.  This procedure is carried out in Appendix I.
\par
The formulation of Eq. 1 is similar to calculations introduced to describe
the anomalous polarizability of $O^{2-}$ in ferroelectrics[23,24], and the
resulting nonlinear waves are also similar to those found in
ferroelectrics[25].
The equation of motion corresponding to Eq. 1 then becomes:
$${M\over\bar R^2}\ddot\phi (\vec l)+{\Gamma_0\over 8}\sum_{\eta}[
cos\phi ({\vec l})(sin\phi ({\vec l})-sin\phi ({\vec l}+\eta a\hat x))^3
-sin\phi ({\vec l})(cos\phi ({\vec l})-cos\phi ({\vec l}+\eta a\hat y))^3]$$
$$+(2\Gamma_0^a-\Gamma_2)sin\phi (\vec l)cos\phi (\vec l)[sin\phi^2({\vec l})
-cos\phi^2({\vec l})]+{2\over\bar R^4}
{\partial H_{epl}\over\partial\phi (\vec l)}=0,\eqno(2)$$
where the last term is given by Eq. A4 in Appendix I.
\bigskip
\noindent{\bf 3b. (110) Waves}
\smallskip\par
In studying the nonlinear wave solutions of Eq. 2, it is convenient to look for
one-dimensional solutions.  The simplest such solutions propagate along the
x or y axes or at 45$^o$.  It is assumed that the background state is of
HTT symmetry (i.e., $sin\phi =cos\phi =0$ except for the modulation produced by
the wave), and that the tilt has the same magnitude for all atoms in any plane
perpendicular to the propagation direction, but with neighboring cages tilting
in the opposite sense.  As suggested by Fig. 1, the most interesting solutions
are those propagating at $45^o$, (110) waves, which include the
LTO-like domain walls, and this case will be analyzed in the present section.
\par
Figure 1 can be used to understand the reduction of the problem to one
dimension.  For convenience, lines of octahedra parallel to the propagation
direction (e.g., A and B in Fig. 1) will be referred to as {\it columns}, and
lines perpendicular to this direction, forming wave fronts, as {\it rows}.
Thus, it can be seen in Fig. 1 that all atoms fall into one of two classes of
columns, A or B, and likewise all rows, and that all of the nearest neighbors
of an A octahedron are B octahedra.  In a one-dimensional, $45^o$ wave, it will
be assumed that all the octahedra within a given row have exactly the same
value of $\phi$, so that the $\phi (\vec l)$ may be rediced to the one
dimensional $\phi_n$, where $n$ denotes the successive row in the propagation
direction.  Because of the antiferrodistortive nature of the transition, the
tilts of adjacent cells tend to be opposite in phase.  It is convenient to
explicitly account for this factor, by defining
$$cos\phi (\vec l)=(-1)^ncos\phi_n.\eqno(3)$$
The problem is thus effectively reduced to one-dimensional.  For the
$45^o$ waves, Eq. 2a takes the form
$${M\over\bar R^2}\ddot\phi_n+{\Gamma_0\over 8}\sum_{\eta}[
cos\phi_n(sin\phi_n-sin\phi_{n+\eta})^3
-sin\phi_n(cos\phi_n-cos\phi_{n+\eta})^3]
-{(2\Gamma_0^a-\Gamma_2)\over 4}sin4\phi_n+
{2\over\bar R^4}{\partial H_{epn}\over\partial\phi_n}=0.\eqno(4)$$
Whereas anharmonic phonons have often been studied as a paradigm for soliton
physics, most models assume a diagonal anharmonicity.  In contrast, the present
model has significant nonlinearity also in the intercell coupling.
\par
Equation 4 has travelling wave solutions of the form:
$$\phi_{n+1}(t+a^{\prime}/v)=\phi_n(t),\eqno(5)$$
where $a^{\prime}=a/\sqrt{2}$ is the effective repeat distance, $a$ is the
Cu-Cu distance and $v$ is the phase velocity.  The continuum limit amounts to
expanding
$$sin\phi_{n\pm 1}(t)=sin\phi_n(t)\pm \tau cos\phi_n(t)\dot\phi_n(t)+{1\over 2}
\tau^2(cos\phi_n(t)\ddot\phi_n(t)-sin\phi_n(t)\dot\phi_n^2(t)),\eqno(6)$$
where $\tau=a^{\prime}/v$, with a similar expansion for $cos\phi$.
Substituting Eq. 6 into Eq. 5 reduces that equation to a single-site form:
$${M\over\bar R^2}\ddot\phi
-{(2\Gamma_0+2\Gamma_0^a-\Gamma_2)\over 4}sin4\phi
+{3\tau^2\Gamma_0\over 8}[\dot\phi^2sin4\phi+2\ddot\phi sin^22\phi ]
+\Gamma_{ep}
[4sin4\phi+\tau^2(\ddot\phi cos4\phi -\dot\phi^2sin4\phi )]=0,\eqno(7a)$$
where terms of higher order than $\tau^2$ have been neglected, and the index
$n$
has been omitted.  The final term of Eq. 7 follows from Eq. A6 of Appendix I,
with
$$\Gamma_{ep}={8\tilde\alpha_-^{e}\over 3\pi\bar R^2}.\eqno(7b)$$
Equation 7 may be rewritten in dimensionless form by introducing
$$\beta^{\prime} ={(16\Gamma_{ep}+\Gamma_2-2\Gamma_0-2\Gamma_0^a)\bar R^2\over
M},\eqno(8a)$$
$$\gamma^{\prime} ={3\tau^2\Gamma_0\bar R^2\over 32M}={v_1^2\over 4v^2},
\eqno(8b)$$
$$\delta^{\prime} ={\tau^2\Gamma_{ep}\bar R^2\over 4M},\eqno(8c)$$
and
$$\alpha =4\phi,\eqno(8d)$$
in which case Eq. 7a becomes
$$\ddot\alpha ={(-\beta +\gamma \dot\alpha^2)sin\alpha\over 1+4\gamma cos
\alpha },\eqno(9)$$
with $\beta =\beta^{\prime}/(1-4\gamma^{\prime})$, and
$\gamma =(\delta^{\prime}-\gamma^{\prime})/(1-4\gamma^{\prime})$.
The quantities $\gamma^{\prime}$ and $\delta^{\prime}$ are intrinsically
positive.  In principle, $\beta^{\prime}$ (Eq. 8a) could be either positive or
negative, but will here be taken positive, to stabilize an LTT-like phase
(potential minima at $\phi =0,$ $\pi /2$, etc.).
However, $\beta$ in Eq. 9 will be negative when $4\gamma^{\prime}
>1$ ($v<v_1$), while $\gamma$ may have either sign.  When the JT coupling
effects dominate over the anharmonicity, then
$\delta^{\prime}>\gamma^{\prime}$,
so $\gamma$ will be positive for fast waves.
The velocity $v_1$ is a critical velocity, which separates fast waves ($v>v_1$)
from slow waves ($v<v_1$).
\bigskip
\noindent{\bf 3c. (100) Waves}
\smallskip\par
The analysis for (100) waves is very similar to the above analysis for (110)
waves.  In this Section, a subscript will be added to various symbols to
distinguish 100 from 110 waves.  Since both waves satisfy the same equation,
the
symbol will be omitted in subsequent sections, unless it is necessary to
distinguish the two types of wave (e.g., in Section 5).  Now the wave
propagates
along the x-direction, with rows along y.
Hence, nearest neighbors along the rows are exactly out of phase,
$$R_1({\vec l}\pm a\hat y)=-R_1({\vec l}).$$
By a calculation similar to that given above, the (100) wave equation has the
same form as Eq. 9, with the following changes.  First, $\tau_{100}=a/v$, since
$a$ is the distance between rows.  Next, since only half of the neighbors now
contribute to the dispersion,
$${\gamma_{100}^{\prime}\over\tau_{100}^2}={\gamma_{110}^{\prime}
\over 2\tau_{110}^2},$$
with a similar relation for $\delta_{100}^{\prime}$.  However, since $a^{\prime
2}=a^2/2$, $\tau_{100}^2=2\tau_{110}^2$, so $\gamma_{100}^{\prime}=\gamma_{110}
^{\prime}$ and $\delta_{100}^{\prime}=\delta_{110}^{\prime}$.  Thus, these two
corrections compensate.  However, there is one last correction, an additional
term in $H_{epn}$, Eq. A9 in Appendix I.  This changes the denominator in
the expressions for both $\beta$ and $\gamma$:
$$(1-4\gamma^{\prime})\rightarrow (1-4(\gamma^{\prime}+\delta^{\prime})).
\eqno(10)$$
Hence, (100) waves have virtually the same dispersion as (110) waves.  Both
waves satisfy Eq. 9, with the only change being the substitution, Eq. 10, for
the (100) waves.  In Section 5, the significance of this difference will become
apparent.  Section 4 analyzes possible solutions of Eq. 9, and hence is
applicable to both types of waves.
\bigskip
\noindent{\bf 4. Nonlinear Wave Solutions}
\smallskip\par
\bigskip
\noindent{\bf 4a. Phase Portraits}
\smallskip\par
A first integral of Eq. 9 can readily be found by setting $y=\dot\alpha$,
$$\ddot\alpha =\dot y={dy\over d\alpha}\dot\alpha =y^{\prime}y,$$
so
$$\int_{y_0}^y{ydy\over \gamma y^2-\beta}=\int_{\alpha_0}^{\alpha}{sin\alpha d
\alpha\over 1+4\gamma cos\alpha},$$
or
$${y^2-\hat\beta\over y_0^2-\hat\beta}=\sqrt{1+4\gamma cos\alpha_0
\over1+4\gamma cos\alpha}\equiv F(\alpha ),\eqno(11)$$
where $\alpha_0$ and $y_0$ are initial values of $\alpha$ and $y$ respectively,
and $\hat\beta =\beta /\gamma$.  Equation 11 may be reduced to quadrature:
$$t=\int_{\alpha_0}^{\alpha}{d\alpha\over\sqrt{F(\alpha )(y_0^2-\hat\beta )+
\hat\beta}}.\eqno(12)$$
This equation must be solved numerically.  Equation 11 depends on two
parameters
$\hat\beta$ and $\gamma$.  In constructing phase portraits, only $\gamma >0$
need be considered, since $\gamma <0$ is equivalent to $\gamma >0$, but with
$\alpha\rightarrow\alpha +\pi$.  The nature of the phase portraits does change,
however, when $\gamma =\gamma_c=1/4$.  Hence, there are four classes of phase
portrait, corresponding to $\hat\beta >,<0$, and $\gamma >,<\gamma_c$.
For LSCO, the fast wave solutions are expected to correspond to $\beta ,\gamma
>0$, and slow waves to $\beta ,\gamma <0$.  The corresponding phase portraits
($\dot\alpha$ vs $\alpha$) are illustrated in Figs. 2 ($\beta ,\gamma >0$), 3
($\beta <0$, $\gamma <0$), and 4 ($\beta <0$, $\gamma >0$).  The lines with
filled circles in these figures are separatrices, dividing regions of
qualitatively different kinds of wave behavior.    The following subsections
will describe the various kinds of waves which can be generated.
\par
The situation corresponding to Fig. 4 does not appear to arise in
LSCO.  However, it could occur under special circumstances, and since the
resulting waves are strikingly different (particularly for $\gamma >\gamma_c$),
solutions for this case will be discussed in Section 4d.
\bigskip
\noindent{\bf 4b. Fast Wave Solutions}
\smallskip\par
For $\beta >0$ and $0<\gamma <\gamma_c$, the typical phase space profile ($\dot
\alpha$ vs
$\alpha$) is shown in Fig. 2a.  This diagram periodically repeats outside the
range illustrated, $\alpha\rightarrow\alpha +2n\pi$, $n$ integer.  The overall
nature of the phase portraits can be understood by noting that, for $\gamma
\rightarrow 0$, Eq. 9 reduces to the equation of a pendulum.  Hence, there will
be low energy solutions corresponding to oscillations about a single potential
minimum, and higher energy solutions in which the octahedron hops between
adjacent minima (the `pendulum' undergoes complete $2\pi$ rotations).  The
finite values for $\gamma$, due to intercell coupling, lead to all the
complications discussed below.
\par
In Fig. 2a, there are two
separatrices, the nature of which can best be understood by looking at a series
of typical orbits, Fig. 5.  These orbits all correspond to the parameters
$\beta
=\gamma =0.2$, $\alpha_0=0$, with different choices of $y_0=\dot\alpha (t=0)$.
This corresponds to moving up the left hand axis of Fig. 2a.  Here and below,
the waveforms are found by numerical integration of Eq. 12.  For the parameters
of Fig. 5, the first separatrix occurs at $y_0=y_1$, with $0.816<y_1<0.817$.
For $y_0<y_1$, Fig. 2a shows that there are $\alpha$ values which are not
sampled (the phase space plot passes through $\dot\alpha =0$).  The resulting
orbit is consequently a bounded periodic orbit.  For $y_0>y_c$, all values of
$\alpha$ are sampled, and, since $\dot\alpha$ never vanishes, $\alpha$
increases
monotonically.
\par
An analytic formula for $y_1$ can readily be found from Eq. 11.  From Fig. 2a,
the separatrix has initial conditions $\alpha_0=0$, $y_0=y_1$, and passes
through the point $\alpha =\pi$, $y=0$.  Then Eq. 11 may be rearranged to
$$y_1^2=\hat\beta\bigl(1-\sqrt{1-4\gamma\over 1+4\gamma}{}\bigr),\eqno(13a)$$
which yields $y_1=0.8165$ for $\beta =\gamma =0.2$.  The time to travel from
$\alpha =0$ to $\alpha =\pi$ is
$$t={1\over\sqrt{\hat\beta}}\int_0^{\pi}\bigl(1-\sqrt{1-4\gamma\over 1+4\gamma
cos\alpha}{}\bigr)^{-1/2}d\alpha.\eqno(13b)$$
When $\alpha =\pi -\alpha^{\prime}$ with $\alpha^{\prime} <<1$, the integrand
is $\sim 1/\alpha^{\prime}$.  Thus, the integral diverges logarithmically: the
separatrix is a `single kink' soliton, varying once from $-\pi$ to $+\pi$ as
$t$ varies from $-\infty$ to $+\infty$.
\par
The second separatrix corresponds to $y_0=y_2\equiv\sqrt{\hat\beta}$.  From Eq.
9, it can be seen that $\ddot\alpha =0$ when $y_0=y_2$.  Hence, $\dot\alpha
=y_2
$ remains constant, and $\alpha$ increases linearly with time.  For $y_0>y_2$,
$\alpha$ again increases monotonically with time.  However, now $\dot\alpha$ is
larger near $\alpha =(2n+1)\pi$, than near $0$, $2n\pi$, so $\alpha$ tends to
linger near the potential minima (lower slope near $2n\pi$).  In contrast, for
$y_0<1$, the orbit tends to avoid the potential minima.  The extreme case is
for
the separatrix $y_0=y_1$, where the orbit is a single kink, Eq. 13.  These
high-$\dot\alpha$ solutions are not of much interest, since
they also correspond to high energies; the low energy solutions are those which
spend most of their time near the potential minima.
\par
Figure 6 shows a number of different periodic orbits, for $y_0<y_1$.  The low
energy waves are the small amplitude, nearly sinusoidal waves which remain
close
to $\alpha =0$.  These correspond to waves in an LTT-type phase: the octahedra
are all tilted in nearly the same x-direction ($\phi =\alpha /4\simeq 0$, but
the tilt axis wanders back and forth as the wave progresses.  As the amplitude
increases, the waves become increasingly nonsinusoidal, tending to spend
more time near the potential maxima $\alpha =\pm\pi$.  These are not of
importance, being high energy waves, but qualitatively similar slow waves are
of
great interest, and are discussed further in Section 4c.
\bigskip
\noindent{\bf 4c. Slow Waves: Soliton Lattices in the `LTO Phase'}
\smallskip\par
When $v$ becomes less than $v_1$, the signs of both $\beta$ and $\gamma$
change.
This sign change is completely compensated for by shifting $\alpha\rightarrow
\alpha +\pi$, as can be seen by comparing Figs. 2 and 4.  Thus, the periodic
orbits of Fig. 6 can be interpreted as fast wave solutions, using the left
hand axis, or equivalently as slow wave solutions (with $\beta$, $\gamma$ of
same magnitude but opposite sign), using the right hand axis.  This simple
switch has a profound effect on the nature of the low energy solutions.  Now
the small amplitude waves are high energy waves, corresponding to small
oscillations about the unstable energy maximum $\phi =\pi /4$.
\par
On the other hand, the large amplitude waves are now low energy solutions,
because they spend most of their time near potential minima at $\phi =0$ or
$\pi
/2$.  These are {\it soliton lattice} solutions, and a particular solution is
displayed in Fig. 7.  Corresponding to the periodic solutions for $y_0<y_1$,
there are unbounded solutions for $y_0>y_1$.  The most interesting of these
are ones, such as illustrated in Fig. 8, which again form soliton lattices.
\par
Since these are travelling wave solutions, Eq. 5, the spatial dependence can be
restored by substituting $t\rightarrow t-x/v$, where in the lattice case,
$x=na^
{\prime}$, with $n$ an integer.  Hence, Fig. 7 can be thought of
as an illustration of the spatial variation of $\phi$ at a fixed instant of
time.  Finally, in the limit of a static solution $v\rightarrow0$, Fig. 7
would be the spatial profile of a static domain array, transforming from
domains of x-tilted LTT phase ($\phi =0$) to domains of y-tilted LTT phase
($\phi =\pi/2$), precisely as in Fig. 1.
\par
The bounded orbit solutions (Figure 7)
may be called `orthons', in that a static array of orthons would produce a
phase of average macroscopic {\it orthorhombic} symmetry -- the material keeps
switching back and forth between an x-tilted domain and a y-tilted domain.
The unbounded solutions (Figure 8) are `tetrons' in that they switch
periodically between domains of tilt $+x$, $+y$, $-x$, $-y$, $+x$, ..., to
produce an average tetragonal symmetry.
\par
The distinction between orthons and tetrons can be more clearly understood by
analyzing the associated strains.  From Ref. [13,14], it is known that the
orthorhombic shear strain $e_{xy}$ is proportional to the product $R_1R_2$,
whereas the LTT phase (in a single layer) is associated with $e_-=e_{11}-e_{22}
\propto R_2^2-R_1^2$.  Figure 7b,c (8b,c) plot the values of $cos2\phi$ and
$sin2\phi$, which should be proportional to $e_-$ and $e_{12}$, respectively,
in
the orthon (tetron) wave.  Within the individual domains, there is a large
value
of $e_-$.  However, the sign of $e_-$ alternates in successive domains, so that
the macroscopic average value of $e_-$ vanishes.  In contrast, the octahedral
shear strain $e_{12}$ is non-zero only within the domain walls.  However, for
an
orthon, {\it this strain has the same sign in every domain wall}, so that there
is a net macroscopic shear $<e_{12}>$, proportional to the domain wall density.
For a tetron, the shear strain is equally likely to be positive or negative.
Thus, despite the presence of many LTT-type domains, the tetron has on average
no macroscopic shear strains, leading to a macroscopic HTT symmetry.
\bigskip
\noindent{\bf 4d. $\gamma >\gamma_c$}
\smallskip\par
\par
When $\gamma >\gamma_c$, Figs. 2b, 3b, 4b, the slope $\dot\alpha$ passes
through
a critical value at special angles $\alpha_c$ such that $cos\alpha_c=1/4\gamma$
(Eq. 11).  For $\hat\beta >0$, $\dot\alpha$ diverges when $y_0>\hat\beta$ (case
A) and $\dot\alpha\rightarrow 0$ when $y_0<\hat\beta$ (case B), leading to
strikingly different behavior.  In case A, the
resulting soliton lattices are qualitatively similar to those found for $\gamma
<\gamma_c$, except that $\alpha$ jumps discontinuously from $\alpha_c$ to $2\pi
-\alpha_c$.  Similar behavior arises when $\hat\beta <0$ (case C), as
illustrated in Fig. 9, for a tetron lattice.  (For case C, orthon arrays with
$\alpha$ discontinuities are also possible.)
This discontinuous jump means that the interface is abrupt, and its properties
must be calculated on a microscopic model, and not in the continuum limit.
\par
For case B, the solutions are again periodic oscillations, similar to those in
Fig. 6.  However, the lines $\alpha =\alpha_c$, $2\pi -\alpha_c$ divide these
oscillations into those centered on $2n\pi$ and those centered on $(2n+1)\pi$.
\par
Soliton arrays with jump discontinuities are an intriguing possibility, but
are found only in case C.  While such a situation does not appear to arise in
LSCO for 100 or 110 waves, it might still occur in other directions or in
materials with different parameters.
Thus, if anharmonic effects were stronger, it would be possible to have $\beta
^{\prime}>0$, but $\delta^{\prime}<\gamma^{\prime}$, so $\hat\beta <0$.
\bigskip
\noindent{\bf 4e. Static Solutions}
\smallskip\par
Some care must be exercised in examining the static limit of Eq. 9. The static
solutions of Eq. 9 correspond to $v\rightarrow 0$, or equivalently
$\tau\rightarrow\infty$; however, the time derivatives are really derivatives
with respect to $t-x/v$.  Hence, changing to spatial derivatives requires
multiplying $\ddot\alpha$ or $\dot\alpha^2$ by $v^2$ before taking the limit.
Hence, the static limit of Eq. 9 still has the form of Eq. 9, but with the
superscript dots now signifying spatial derivatives, with $\beta\rightarrow
-\beta^{\prime}/4\gamma^{\prime}v^2$ and $\gamma\rightarrow
-(\delta^{\prime}-\gamma^{\prime})/4\gamma^{\prime}$.  That is, both $\beta$
and
$\gamma$ are expected to be negative in this limit.  In this case, the phase
profiles are identical to the slow waves, discussed in Section 4c.
\bigskip
\noindent{\bf 4f. Energy of a Soliton}
\smallskip\par
While the above analysis shows what kinds of nonlinear waves are {\it
possible},
no account was taken as to the energy of the wave, to
see which waves are {\it probable}.  The low energy waves will be those which
are close in energy to the ground state of the system -- i.e., to the LTT phase
with $\phi$ a multiple of $\pi /2$.  These low-energy waves can be of two
forms.
First, there are ordinary phasons -- nearly sinusoidal excitations of small
amplitude away from the LTT phase, as in Fig. 6.
\par
More interesting, however, are the soliton lattice solutions, Figs. 7-9.  These
will be low energy states when the steps are flat topped and lie close to
an LTT value, $\phi =n\pi /2$, for integral $n$.  Orthon solutions of this form
can be generated from Eq. 11 by chosing $y_0=0$ and $\alpha_0$ small.  As
$\alpha_0$ gets progressively smaller, the steps become flatter and the
periodicity increases.  In the limit $\alpha_0\rightarrow 0$, the solution
approaches the separatrix, which is an isolated soliton.  In the opposite
limit,
choosing $\alpha_0 =0$ and $y_0$ small generates low energy tetron arrays.
\par
The energy per unit cell associated with the wave can be found from Eq. 1.
Applying the same transformations as to the equation of motion, the
energy is found to be
$$E={\tilde M\bar R^2\over 2}\dot\phi^2-(\beta_e-\gamma_e\dot\phi^2)cos4\phi+
\delta_e\ddot\phi sin4\phi,\eqno(14)$$
with
$$\beta_e=\bigl({8\Gamma_{ep}+\Gamma_2-2\Gamma_0^a-2\Gamma_0\over 16}\bigr)\bar
R^4,\eqno(15a)$$
$$\gamma_e={(4\Gamma_{ep}-5\Gamma_0)\bar R^4\tau^2\over 16},\eqno(15b)$$
$$\delta_e={(\Gamma_{ep}-\Gamma_0)\bar R^4\tau^2\over 8},\eqno(15c)$$
$$\tilde M=M-{3\Gamma_0\bar R^2\tau^2\over 8}.\eqno(15d)$$
\par
The static limit of Eq. 14 follows from the substitution
$$\tau{d\over dt}\rightarrow -{d\over d\xi},\eqno(16)$$
with $\xi =x/a^{\prime}$ (for (110) waves).  Denoting the spatial derivative on
the right side of Eq. 16 by a prime, Eq. 14 becomes
$$E=E_0\bigl(-{3\over 8}\phi^{\prime 2}-(\hat\beta_e-\hat\gamma_e\phi^{\prime
2})cos4\phi+\hat\delta_e\phi^{\prime\prime} sin4\phi\bigr),\eqno(17)$$
with
$$E_0=\Gamma_0\bar R^4,\eqno(18a)$$
$$\hat\beta_e={8\eta_{ep}+\eta_2-2\over 16},\eqno(18b)$$
$$\hat\gamma_e={4\eta_{ep}-5\over 16},\eqno(18c)$$
$$\hat\delta_e={\eta_{ep}-1\over 8},\eqno(18d)$$
and $\eta_{ep}=\Gamma_{ep}/\Gamma_0$, $\eta_2=(\Gamma_2-2\Gamma_0^a)/\Gamma_0$.
The static limit of the parameters of Eq. 9 can also be
expressed in terms of the $\eta$ parameters.  However, these differ for 110 and
100 waves:
$$\beta_{110} =-{8\over 3}(16\eta_{ep}+\eta_2-2),\eqno(19a)$$
$$\gamma_{110} =-{1\over 12}(8\eta_{ep}-3).\eqno(19b)$$
$$\beta_{100} =-8\Bigl({16\eta_{ep}+\eta_2-2\over 8\eta_{ep}+3}\Bigr),
\eqno(19c)$$
$$\gamma_{100} =-{1\over 4}\Bigl({8\eta_{ep}-3\over 8\eta_{ep}+3}\Bigr).
\eqno(19d)$$
Note that $\hat\beta$ remains the same for both kinds of waves.
\par
Figure 10 illustrates the energy per unit cell, Eq. 17, for the oscillatory
sloutions of Fig. 6, assuming $\eta_{ep}=2$, $\eta_2=0$. [N.B.: This is only
done for illustrative purposes.  Strictly speaking, the assumed $\eta$ values
are not consistent with the parameters of Fig. 6; this will be corrected in
Section 5, after the parameters relevant to LSCO have been estimated.]
For simplicity, the energy associated with only one half-cycle of oscillation
(i.e., a single soliton) is explicitly displayed.  In this case, the
soliton has a simple lineshape, the energy being dominated by the term $\hat
\beta_e$ (Eq. 18b).  The linewidth is $\sim 5a/\sqrt{\hat\beta}$, and the
profile quickly becomes independent of soliton spacing -- i.e., there is
very little soliton-soliton interaction in the present model.
\bigskip
\noindent{\bf 5. Applications to LSCO}
\smallskip\par
Numerical estimates for the above parameters follow from the calculations of
Ref. [13].  These calculations may underestimate the parameters, since it was
assumed that the HTT$\rightarrow$LTO transition represents the onset of
tilting,
$\bar R\ne 0$, whereas it is more probable that $\bar R$ is finite in the HTT
phase, but with $<cos\phi >$=0.  Furthermore, the value of $\bar R$ is taken
from the measured tilt, whereas this macroscopic average value may only be
about half as large as the instantaneous microscopic value[14].  With these
caveats, the best estimates are from parameter sets 10 and 11 of Table II of
Ref. [13].  The resulting parameter values are summarized in Table I.  The
Table
also includes values for the crossover velocity, $v_1$, where
$$v_1^2={3E_0a^2\over 16M\bar R^2}.$$
\par
Note the striking difference of $\gamma$ values for the two types of waves.
For
110 waves, $\gamma$ is large and negative, dominated by the electron-phonon
coupling.  However, for 100 waves, $|\gamma |$ is always restricted to be less
than $\gamma_c=1/4$.  Hence, the slow wave and static solutions for 100 waves
are described by the phase profiles of Fig. 3a, and include orthons, tetrons,
and isolated kinks.  By contrast, the 110 waves correspond to Fig. 3b, so the
low energy solutions are phasons:
weak periodic modulations of the tilt axis about some LTT state.
\par
Figure 11 illustrates the energy of a soliton wall, using Eqs. 18 and 19 to
estimate all parameters, and using typical values $\eta_{ep}=8$, $\eta_2=0$,
and $E_0=5.5meV$.  Shown are both the total energy (solid line) and the various
components (corresponding to Eqs. 18a-d).  The shape is more complicated than
found in Fig. 10, because the term in $\hat\gamma_e$ (Eq. 18c) becomes
comparable to that in $\hat\beta_e$ when $|\hat\beta |$ is large.
\par
By integrating the energy profile, the total interface energy $E_w$ (plotted as
energy per column width) can be found, and is plotted in Fig. 12 as a function
of the spacing between successive domain walls, $d_w$.  In the present theory,
the tilt-tilt interaction is short ranged, and this is reflected in Fig. 12 by
the fact that there is virtually no soliton-soliton interaction until the
solitons are less than two cells apart.  Below this separation, the tetrons
have
lower energy than the orthons.  However, it is unlikely that a continuum theory
should be trusted at such small length scales -- note that the width of the
domain wall is less than a single cell across.
The limiting energy of a single
soliton at large separation is found to be nearly linear in $\eta_{ep}$:
$$E_{w}\simeq{E_0\over 2}(\eta_{ep}-1.3).\eqno(20)$$
\par
There are a number of additional factors which will give rise to longer range
soliton interaction.  For instance, the strains associated with the tilting
give rise to direct coupling of more distant cells[14].  Moreover, the
electronic energy lowering is produced by the splitting of the vHs degeneracy
within the LTT-like domains. If the domain walls approach one another too
closely, the domains will cease to have a well-defined LTT-like character, and
the vHs splitting will decrease.  Thie is equivalent to a repulsive interaction
between domain walls.
\par
Using the present equations, it is possible to calculate the energy associated
with a static domain wall exactly, for a given microscopic configuration, Fig.
13.  Such calculations are presented in Appendix II.  These calculations are
still approximate, in that the simple distortions assumed do not display the
correct corner sharing behavior.  The resulting energy is very similar
to the calculations of this section.  The wall energy is found to satisfy an
equation of the form of Eq. 20, but with different numerical coefficients,
which depend explicitly on the assumed wall thickness.  The minimum wall energy
corresponds to an abrupt wall, as in Fig. 13a.
\par
It is tempting to relate the various nonlinear waves to the various phases
found
in LSCO and related compounds -- the isolated solitons separating large grains
of LTT phase, the orthons associated with the LTO phase, and the tetrons with
the HTT phase.  If this correspondence is valid, then what distinguishes
the various phases, and in particular, what constitutes a phase transition?
Presumably, this is accomplished by coupling to macroscopic strains:  e.g., in
the LTO phase, the macroscopic average shear strain, $\epsilon_{xy}$.  As shown
in Ref. [14], inclusion of strain coupling makes the effective tilt-tilt
interaction longer ranged, and hence could stabilize an orthon or tetron phase.
[This same model could be applied to BaTiO$_3$, in which case, the long-range
coupling would be via the ferroelectric polarization.]
\bigskip
\noindent{\bf 6. Charged Solitons?}
\smallskip\par
\par
In the above sections, it has been shown that the LTO phase of LSCO can be
interpreted as a dynamic JT phase, or equivalently as a soliton lattice with
domains of LTT phase separated by domain walls of a {\it different phase},
closer to the usual LTO phase.  This nanoscale phase heterogeneity is very
reminiscent of another grain boundary phase postulated[7,26] to exist in
this system -- a phase separation of {\it holes} associated with doping LSCO
away from the optimal hole concentration at which the vHs coincides with the
Fermi level.
\par
In this earlier model, the phase separation arises because the electron-phonon
coupling is strongest exactly at the vHs, leading to a minimum in the free
energy at precisely that concentration.  (There is a close relation to
Hume-Rothery phases in alloys.)  The size of the domains is restricted, because
the hole concentration is different in the two phases -- i.e., the domains are
charged, and the domain size is limited by Coulomb effects.  The lowest energy
phase turns out to be a grain boundary phase, with the LTT domains forming one
phase, and the domain walls constituting a second, microscopic LTO-like phase.
Near the vHs, it is possible to have a pure LTT phase.  As the material is
doped
away from the vHs toward half-filling, the LTO phase becomes more stable, and
the size of the LTT domains decreases.  If, as found above (Fig. 7), the
orthorhombic shear strain is taken as proportional to the domain wall density,
this is exactly what is found experimentally in LSCO -- that $e_{12}$ increases
as the material is doped away from the vHs ($x$ decreased below $\sim 0.15$).
\par
There are a number of similarities between the doping-induced grain boundary
phase and the soliton lattice.  Indeed, it is possible that the solitons are
charged.  This could arise if the strains in the domain wall, associated with
corner sharing, lead to a modification of the {\it shape} of the octahedron --
i.e., a change of the Cu to apical O bond length.  A reduction of this bond
length would reduce the splitting between the Cu $d_{3z^2-r^2}$ and $d_{x^2-
y^2}$ levels, which can lead to a different degree of
charge transfer into the Cu-O$_2$ planes, as well as to a direct change in
the shape of the Fermi surface.  These changes will have similar effects to the
charge bunching postulated in the earlier work[26], in that the density of
planar Cu holes will be different in the domains and in the walls.  It is
of interest to note that a number of experiments have detected structural
anomalies in the cuprates, often at temperatures very close to T$_c$[17], which
are related to tilting of the octahedra, and generally include a change of the
Cu to apical O bond length.
\bigskip
\noindent{\bf 7. Conclusions and Discussion}
\smallskip\par
The present paper is the fourth in a series of closely related works which have
introduced the idea that the LTO phase of LSCO is a dynamic JT phase.  It is
perhaps appropriate to briefly summarize what has been accomplished to date,
and
what remains to be done.  The formal model was introduced in B[13], and static
JT solutions were displayed, emphasizing the first appearance of a macroscopic
tilt, $\bar R\ne 0$.  The dynamics of the model, wherein the tilt axis could
bob about, assuming one of four preferred orientations, is discussed in C[14]
at
the mean-field level, and in D, the present manuscript, in terms of a
microscopic picture of nonlinear wave solutions.  The first paper, A[12],
introduced the concept of the dynamic vHs-JT phase, and provided a simplified
model calculation which reproduced the general doping dependence of the phase
diagram.  In particular, it answered the vexing question of why, if
superconductivity can only be found in the LTO phase[11], the highest
superconducting transition temperatures occur in the doping range in which the
LTO phase is least stable.
\par
Prior to these papers, the vHs model seemed to provide an interesting framework
for interpreting many of the phenomena associated with high-T$_c$
superconductivity[7,27], but also predicted a strong competition with
structural instabilities (charge density wave phenomena) of a form which could
not be clearly identified with the major phase transitions of the high-T$_c$
cuprates.  While the discovery of the LTT phase provided a first link between
structural instability and the vHs, the present model clearly demonstrates for
the first time a close relation between the vHs and the LTO phase, the dominant
phase in which high-T$_c$ superconductivity arises.  Moreover, the model paints
this LTO phase as a highly unusual, dynamic and nonlinear phase, of just the
form wherein one might expect to observe the anomalous normal-state and
superconducting properties actually found in the cuprates.  The close
similarity between the present model and the earlier picture of short-range
structural disorder[7,26] should be noted.
\par
While the present equations display a rich variety of nonlinear wave solutions,
it should be kept in mind that these equations are already approximations, and
that the more exact equations[13] should have even stronger nonlinearities.  In
particular, it was assumed that the tilting of the individual octahedral cages
sets in at considerably higher temperatures, so that the average tilt $\bar
R$ can be considered constant.  Moreover, while the term $H_{ep}$ in Eq. 1
accounts for the role of the electronic excitations in modifying the structural
transition, the back-reaction of this structural change on the electronic
properties has not been discussed.  In particular, what is the effect of the
domain walls on the splitting of the vHs?  Indeed, in a dynamic JT phase, it
is unclear what the `band structure' means, when the Brillouin zone itself may
be a local function of time and space (c.f. the spatial variation of the
orthorhombic splitting in Figs. 7 and 8).
\par
It should be noted that these complications are `interesting' in that they are
likely to lead to a rich variety of anomalous behavior, both in the
superconducting and normal states, comparable to that observed experimentally
in
the cuprates.
\par
{\bf Acknowledgments:} I would like to thank J. Jos\'e, A. Widom, and F.Y. Wu
for stimulating conversations.  Publication 546 from the Barnett Institute.
\bigskip
\centerline{\bf Appendix I: Approximate Electron-Phonon Coupling Term}
\bigskip\par
In this Appendix, an approximate form for the electron-phonon coupling term,
Eq.
1b, will be derived. Following the discussion of Section 3, this term can be
written
$$H_{epn}=-{\tilde\alpha_-^{e}\bar R^2\over 4}|
\sum_{\eta}[(cos\phi_n-cos\phi_{n+\eta})^2-(sin\phi_n-sin\phi_{n+\eta}))^2]|.
\eqno(A1)$$
Expanding $sin\phi_{n+\eta}$, $cos\phi_{n+\eta}$ as in Eq. 6, this becomes
$$H_{epn}=-{\tilde\alpha_-^{e}\bar R^2\over 2}|
4cos2\phi-\tau^2(2\ddot\phi sin2\phi +3\dot\phi^2cos2\phi )|,\eqno(A2a)$$
$$H_{epn}=-{\tilde\alpha_-^{e}\bar R^2\over 2}(4-3\tau^2\dot\phi^2)
\sqrt{1+\mu^2}|cos(2\phi+\nu )|,\eqno(A2b)$$
where the index $n$ on $\phi$ is omitted, $\mu =2\tau^2\ddot\phi
/(4-3\tau^2\dot
\phi^2)$, and $sin\nu =\mu /\sqrt{1+\mu^2}$.  Figure 14a illustrates $H_{epn}$
in the case of perfect intercell coupling, $\cos\phi_{n\pm 1}=-cos\phi_n$
($\dot
\phi =\ddot\phi =0$).  Deviations from this coupling reduce the amplitude of
$H_
{epn}$ and shift the phase of $cos2\phi$.
\par
The absolute value signs in Eq. A2 lead to the cusps in Fig. 14a, and can
complicate the analysis.  Since $H_{epn}$ is an effective potential for the
tilt phonons, rounding the cusps at the potential {\it maxima} should have
negligible effect on the problem.  Hence, the Fourier expansion may be
utilized:
$$|cos2\phi |={2\over\pi}(1+\sum_{n=1}^{\infty}{2(-1)^{n+1}\over 4n^2-1}cos4n
\phi ).\eqno(A3)$$
Keeping only the constant and $n=1$ terms provides a reasonable first
approximation (dashed line in Fig. 14a).
\par
What enters into the equations of motion is not $H_{epn}$ but its derivative
with respect to $\phi_n$,  which may be written
$${\partial H_{epn}\over\partial\phi_n}=
(\pm ){\tilde\alpha_-^{e}\bar R^2\over 2}
[4sin2\phi+\tau^2(\ddot\phi cos2\phi -\dot\phi^2sin2\phi )],\eqno(A4a)$$
$$=(\pm ){\tilde\alpha_-^{e}\bar R^2\over 2}(4-\tau^2\dot\phi^2)
\sqrt{1+\mu^{\prime 2}}sin(2\phi+\nu^{\prime}),\eqno(A4b)$$
where, to order $\tau^2$, $\mu^{\prime}=\mu$ and $\nu^{\prime}=\nu$, and the
term $(\pm )$ comes from the absolute value signs in Eq. A1: the minus sign
applies whenever the quantity within $||$'s in Eq. A1 is negative.  The
resulting quantity, plotted in Fig. 14b, has discontinuous jumps whenever
the $cos2\phi$-term changes sign, Fig. 14a.  Because of these discontinuities,
the Fourier series requires more terms for an adequate representation:
$$(\pm )sin2\phi ={8\over\pi}\sum_{n=1}^{\infty}{n(-1)^{n+1}\over 4n^2-1}sin
4n\phi.\eqno(A5)$$
Figure 14b shows the Fourier series for one term (dashed line) and for two
(dotted line).  The one term series is nonetheless qualitatively correct,
particularly in the region of the potential minima, so this approximation is
adopted for the main text.  In the present case, this amounts to replacing
$$(\pm )sin(2\phi +\nu)\rightarrow{8\over 3\pi}sin(4n\phi +\nu),\eqno(A6)$$
so
$${\partial H_{epn}\over\partial\phi_n}=
{4\tilde\alpha_-^{e}\bar R^2\over 3\pi}
[4sin4\phi+\tau^2(\ddot\phi cos4\phi -\dot\phi^2sin4\phi )].\eqno(A7)$$
\par
There is one last complication.  The terms $H_{ep,n\pm 1}$ also depend on $\phi
_n$, so their derivatives must be added to Eq. A7 to give the total potential
contribution.  By reasoning similar to the above, it can be shown that
$${\partial H_{ep,n+1}\over\partial\phi_n}+
{\partial H_{ep,n-1}\over\partial\phi_n}=
{\partial H_{epn}\over\partial\phi_n},\eqno(A8)$$
except possibly when $cos2\phi_n$ is changing sign.  In this case, it is
possible that, e.g., ${\partial H_{epn+1} /\partial\phi_n}$ will change sign
before ${\partial H_{epn} /\partial\phi_n}$ (Fig. 14b).  In this case, its
contribution would tend to cancel rather than add.  This effect will tend to
smooth out the discontinuity in Fig. 14b.  Furthermore, the correction will be
small in the smoothed form of the restoring force (dashed line in Fig. 14b).
Hence, this complication is neglected, and Eq. A8 is assumed to hold
throughout.
This amounts to doubling the contribution of $H_{epn}$, accounting for a factor
of 2 in Eq. 2 and subsequent equations.
\par
Finally, the above analysis can be repeated for (100) waves, as in Section 3c.
In this case, Eq. A7 becomes
$${\partial H_{epn}\over\partial\phi_n}=
{4\tilde\alpha_-^{e}\bar R^2\over 3\pi}
[4sin4\phi+{\tau^2\over 2}(\ddot\phi (cos4\phi +1)-\dot\phi^2sin4\phi )].
\eqno(A9)$$
Note that this differs from Eq. A7, in that the $\tau^2$ terms are half as
large, and there is an extra term multiplying $\ddot\phi$.
\bigskip
\centerline{\bf Appendix II: Microscopic Domain Wall Energy Calculations}
\bigskip\par
For narrow domain walls (only a few unit cells thick), it is possible to
directly calculate the domain wall energy from Eq. 1, for any assumed pattern
of
tilt angles.  In this Appendix, the
calculation is presented for the three configurations illustrated in Fig. 13.
In this figure, the dashed lines enclose the regions of canted cells --
everything outside the dashed lines is assumed to be in perfect LTT order.
(For the abrupt wall of Fig. 13a, there are no cells `within' the wall.)
The domain wall energy is defined as the excess energy over the ground state
energy within a single domain.  For an LTT-type domain, e.g., $cos\phi =0$
everywhere, this minimum energy is, for a single cell,
$$E_{min}=\bigl({\Gamma_0+\Gamma_0^a\over 2}\bigr)\bar R^4
-2\tilde\alpha_-^e\bar R^2.\eqno(B1)$$
(Recall that the terms involving $\Gamma_0^a$ and $\Gamma_2$ are on-site terms,
while the terms in $\Gamma_0$ and $\tilde\alpha_-^e$ involve sums over nearest
neighbors.)
\par
For the abrupt wall of Fig. 13a, all atoms remain tilted about the x or y axes,
so there is no wall energy contribution from the terms involving $\Gamma_2$ or
$\Gamma_0^a$.  Here, y is defined as parallel to the domain wall and x
perpendicular to it, with the positive x direction pointing to the right of the
figure.  The terms in $\Gamma_0$ and $\tilde\alpha
_-^e$ are affected by the wall, since the cells nearest the walls have
neighbors in a less than optimal configuration.  For instance, for the top left
atom in the figure, the excess energy due to $\Gamma_0$ is
$$\Delta E_{\Gamma}={\Gamma_0\bar R^4\over 4}(2\times 1+0+({1\over 2})^4-2).$$
Here the first term in parentheses is due to the term involving nearest
neighbors along y, the next to the neighbor along --x, the third along +x
(i.e.,
across the domain wall); the final term is the subtraction of the energy for
the
undistorted state (absence of the domain wall).  For the top right atom, the
analogous term is
$$\Delta E_{\Gamma}={\Gamma_0\bar R^4\over 4}(2\times 0+1+({1\over 2})^4-2).$$
For the electron-phonon term, taken from Eq. A1, the net wall energy due to
both
atoms is
$$\Delta E_{\alpha}=-\tilde\alpha_-^e\bar R^2(|2\times 1-({1\over 2})^2|+|1+
({1\over 2})^2-0|-4).$$
Adding up the excess energies of these two cells, the wall energy is found to
be
$$E_w^a=\tilde\alpha_-^e\bar R^2-\bigl({7\over 32}\bigr)\Gamma_0\bar R^4$$
$$=\Bigl({3\pi\over 8}\eta_{ep}-{7\over 32}\Bigr)E_0=1.18E_0(\eta_{ep}-0.186).
\eqno(B2)$$
\par
This procedure can be repeated for the thicker walls of Figs. 13b, 13c.  In
each case, the sum involves one additional cell, and there is now an internal
degree of freedom, the tilt angle $\theta$ of the cells within the wall.  For
the one-cell wall, Fig. 13b,
$$E_w^b={E_0\over 4}\Bigl(2cos^4\theta-3+{sin^4\theta +(1+sin\theta )^4\over 8}
+{sin^22\theta\over 2}\eta_2+({3\pi\eta_{ep}\over 8})[11-2sin\theta-|10cos^2
\theta-3-2sin\theta|]\bigr)$$
$$=2.65E_0(\eta_{ep}+0.047\eta_2-0.133),\eqno(B3)$$
where the final numerical form is for the symmetric wall position, $\theta =45^
o$.  Actually, the term in $\eta_{ep}$ is minimized at $\theta=0$, and so
dominates the wall energy that the total $E_w^b$ also has a minimum at $\theta
=
0$ as long as $\eta_{ep}\ge 1$.  In fact, this limiting case of $\theta =0$
corresponds exactly to the abrupt wall situation, Fig. 13a.  It is a check of
the calculation that Eq. B3 agrees with Eq. B2 in this limit.
\par
The same calculation may be repeated for Fig. 13c.  The wall energy is
$$E_w^c={E_0\over 4}\Bigl(-3-sin^22\theta +{sin^4\theta +(sin\theta +cos\theta
)^4+(1+cos\theta )^4\over 8}+sin^22\theta\eta_2+{3\pi\eta_{ep}\over 4}[9-2cos
\theta -5cos2\theta ]\Bigr)$$
$$=2.80E_0(\eta_{ep}+0.067\eta_2-0.160),\eqno(B4)$$
with the last line holding for $\theta =30^o$.  No absolute value signs appear
in Eq. B4, since it only makes sense for $\theta\le 45^o$ (otherwise the wall
tilt is no longer monotonic).  Again, the wall energy actually minimizes at
$\theta =0$, in which case it agrees with Eq. B2.  Thus, the minimum wall
energy corresponds to the abrupt wall, Fig. 13a.  The energy, Eq. B2, should
be compared to the continuum theory, Eq. 20.  It is seen that the continuum
theory underestimates the absolute energy by a factor of about two, but
correctly predicts that the wall thickness is of the order of a single unit
cell.  It should be noted that the calculations of this Appendix employ the
full, discontinuous form of the electron-tilt coupling (solid line in Fig. 14a)
rather than the smooth approximation (dashed line).
\par
Note that the present model does not fully account for the complications of
corner-sharing octahedra.  The distortion of, e.g., Fig. 13a is not consistent
with a tilting of rigid octahedra.
For instance, in Fig. 13a, the top atom to the left of the
domain wall is tilting parallel to the wall (the upper apical oxygen is tipped
in the positive y direction), so the in-plane oxygen in the right corner of the
cell (positive x direction) would be untilted, for a rigid octahedron.  On the
other hand, the apical oxygen of the top cell to the right of the wall is
tipped
in the negative x direction, which would require the same in-plane oxygen to
tip below the plane.  In the actual corner-shared configuration, neither tilt
can be a pure x or y directed tilt.  This distortion in turn will cause cells
further from the domain wall to be tilted, and may lead to distortions of the
{\it shape} of the octahedra.
\par
However, a more realistic model, allowing for these distortions, would be
considerably more involved.  For instance, it would include variations in the
Cu-apical O
distance, with the ensuing relative change of the Cu $d_{3z^2-r^2}-d_{x^2-y^2}$
level splitting.  While these effects may play an important role in the
detailed
characterization of the domain wall and its excitations, it is hoped that the
present simplified model captures the essential nature of the structural
nonlinearities of the model.
\bigskip
\hrule
\bigskip
\centerline{\bf Table I: Parameter Values (from Ref. [13])}
\vskip 0.25 in
\settabs
\+fitvv&$\omega_e$ vvvv&$i\omega_0$ vvvv&$\omega_{\Gamma}$ vvvv&
$g_1vvvvv$&$g_2$vvvvv&$g_4$vvvvv&$\epsilon_1(0K)$ vv&$
\epsilon_2(0K)$ vv&$\epsilon_1(700K)$ v&$Z_0(0K)$ \cr 
\+fit&$E_0$ &$\eta_{ep}$ &$\eta_2$&$\hat\beta$&$\gamma_{110}$&$\gamma_{100}$&
$v_1$&
\cr
\+&(meV)&&&&&&(m/s)&\cr
\vskip 0.2 in
\+10&6.4&5.6&0.7&68&-3.5&-0.22&470\cr
\smallskip
\+11&4.7&9.6&-2.0&65&-6.2&-0.23&400\cr
\smallskip
\hrule
\vskip 0.2 in
\bigskip
\centerline{\bf References}
\bigskip
\smallskip
\item{[1]}C.A. Balseiro and L.M. Falicov, Phys. Rev. B{\bf 20}, 4457 (1979).
\smallskip
\item{[2]}J.E. Hirsch and D.J. Scalapino, Phys. Rev. Lett. {\bf 56}, 2732
(1986).
\smallskip
\item{[3]}J.D. Jorgensen, H.-B. Schuttler, D.C. Hinks, D.W. Capone, II, K.
Zhang, M.B. Brodsky, and D.J. Scalapino, Phys. Rev. Lett. {\bf 58}, 1024
(1987);
J. Labbe and J. Bok, Europhys. Lett. {\bf 3}, 1225 (1987); Y.E. Dzyaloshinskii,
Pis'ma Zh. Eksp. Teor. Fiz. {\bf 46}, 97 (1987) [JETP Lett. {\bf 46}, 118
(1987).
\smallskip
\item{[4]}J.P. Pouget, C. Noguera, and R. Moret, J. Phys. (France) {\bf 49},
375 (1988).
\smallskip
\item{[5]}P.C. Pattnaik, C.L. Kane, D.M. Newns, and C.C. Tsuei, Phys. Rev.
B{\bf
45}, 5714 (1992).
\smallskip
\item{[6]}J.C. Phillips, Phys. Rev. B{\bf 45}, 12647 (1992).
\smallskip
\item{[7]}R.S. Markiewicz, Int. J. Mod. Phys. B{\bf 5}, 2037 (1991).
\smallskip
\item{[8]}J.D. Axe, A.H. Moudden, D. Hohlwein, D.E. Cox, K.M. Mohanty, A.R.
Moodenbaugh, and Y. Xu, Phys. Rev. Lett. {\bf 62}, 2751 (1989); T. Suzuki
and T. Fujita, J. Phys. Soc. Japan {\bf 58}, 1883 (1990);
M. Sera, Y. Ando, S. Kondoh, K. Fukuda, M. Sato, I. Watanabe, S.
Nakashima, and K. Kumagai, Sol. St. Commun. {\bf 69}, 851 (1989).
\smallskip
\item{[9]}Y. Koike, T. Kawaguchi, N. Watanabe, T. Noji, and Y. Saito, Sol. St.
Commun. {\bf 79}, 155 (1991);
Y. Maeno, N. Kakehi, M. Kato, and T. Fujita, to be published, Phys.
Rev. B., and Physica C{\bf 185-9}, 909 (1991); M. Sato, N. Sera, S. Shamoto,
and S. Yamagata, Physica C{\bf 185-9}, 905 (1991).
\smallskip
\item{[10]}R.S. Markiewicz, J. Phys. Condens. Matt. {\bf 2}, 6223 (1990);
S. Barisic and J. Zelenko, Sol. St. Commun. {\bf 74}, 367 (1990);
W.E. Pickett, R.E. Cohen, and H. Krakauer, Phys. Rev. Lett. {\bf 67},
228 (1991).
\smallskip
\item{[11]}H. Tagaki, R.J. Cava, M. Marezio, B. Batlogg, J.J. Krajewski, W.F.
Peck, Jr., P. Bordet, and D.E. Cox, Phys. Rev. Lett. {\bf 68}, 3777 (1992).
\smallskip
\item{[12]}R.S. Markiewicz, Physica C{\bf 193}, 323 (1992).
\smallskip
\item{[13]}R.S. Markiewicz, to be published, Physica C.
\smallskip
\item{[14]}R.S. Markiewicz, unpublished.
\smallskip
\item{[15]}J. Labb\'e and J. Friedel, J. Phys. (Paris) {\bf 27}, 153, 303,
708 (1966).
\smallskip
\item{[16]}E. Pytte, Phys. Rev. Lett. {\bf 25}, 1176 (1970); Phys. Rev B{\bf
4}, 1094 (1971).
\smallskip
\item{[17]}B.H. Toby, T. Egami, J.D. Jorgensen, and M.A. Subramanian,
Phys. Rev. Lett. {\bf 64}, 2414 (1990); T. Egami, H.D. Rosenfeld, and B.H.
Toby,
Ferroelectrics {\bf 120}, 11 (1991).
\smallskip
\item{[18]}E.K.H. Salje, ``Phase Transitions in Ferroelastic and Co-elastic
Crystals" (Cambridge, University Press, 1990); J.A. Gonzalo, ``Effective Field
Approach to Phase Transitions and Some Applications to Ferroelectrics"
(Singapore, World, 1991), p. 74.
\smallskip
\item{[19]}K.A. M\"uller, in ``Nonlinearity in Condensed Matter", ed. by A.R.
Bishop, D.K. Campbell, P. Kumar, and S.E. Trullinger (Berlin, Springer, 1987),
p. 234.
\smallskip
\item{[20]}R. Comes, M. Lambert, and A. Guinier, Sol. St. Commun. {\bf 6}, 715
(1968).
\smallskip
\item{[21]}Y. Koyama and Y. Hasebe, Phys. Rev. B{\bf 36}, 7256 (1987), B{\bf
37}, 5831 (1988); Y. Koyama and H. Hoshiya, Phys. Rev. B{\bf 39}, 7336 (1989).
\smallskip
\item{[22]}J.C.M. Tindemans -- van Eijndhoven and C.J. Kroese, J. Phys. C{\bf
8}, 3963 (1975); K.-H. H\"ock, G. Schr\"oder, and H. Thomas, Z. Phys. B{\bf
30},
403 (1978).
\smallskip
\item{[23]}H. Bilz, G. Benedek, and A. Bussmann-Holder, Phys. Rev. B{\bf 35},
4840 (1987).
\smallskip
\item{[24]} A. Bussmann-Holder, A. Simon, and H. B\"uttner, Phys. Rev. B{\bf
39}, 207 (1989).
\smallskip
\item{[25]}G. Benedek, A. Bussmann-Holder, and H. Bilz, Phys. Rev. B{\bf 36},
630 (1987).
\smallskip
\item{[26]}R.S. Markiewicz, J. Phys. Condens. Matt., {\bf 2}, 665 (1990).
\smallskip
\item{[27]}J. Friedel, J. Phys. Cond. Matt. {\bf1}, 7757 (1989); D.M. Newns,
C.C. Tsuei, P.C. Pattnaik, and C.L. Kane, Comments in Cond. Matt. Phys. {\bf
15}, 273 (1992).
\bigskip
\centerline{\bf Figure Captions}
\bigskip
\item{Fig.~1}Proposed domain wall between two LTT domains, for (a) (110)
wall or (b) (100) wall.  Circles represent
apical oxygens in layer above CuO$_2$ plane.  Arrow indicates direction of
displacement of apical oxygen due to tilting of corresponding octahedron.
Dashed lines indicate limits of domain wall, within which tilt direction is
variable.  Dotted lines indicate two columns of inequivalent atoms, $A$ and
$B$.
\smallskip
\item{Fig.~2}Phase portrait of Eq. 9 for fast waves ($v>v_1$), with $\beta >0$,
$\gamma >0$.  Lines with filled circles indicate separatrices.
\smallskip
\item{Fig.~3}Phase portrait of Eq. 9 for $\beta <0$, $\gamma <0$.
\smallskip
\item{Fig.~4}Phase portrait of Eq. 9 for $\beta <0$, $\gamma >0$.
\smallskip
\item{Fig.~5}Temporal (or spatial) evolution of fast waves (corresponding to
Fig. 2), assuming $\beta =\gamma =0.2$, $\alpha_0=0$, and $y_0$ = 2, 1, 0.82,
and 0.8, in order of decreasing slope; dashed line is $y_0=2$ curve, with
expanded horizontal scale (upper axis).
\smallskip
\item{Fig.~6}Oscillatory wave solutions, which can equally correspond to
either fast waves (using left-hand axis) or slow waves (right-hand axis).
Parameters (for slow waves) are $\beta =\gamma =-0.1$, $y_0=0$, and $\alpha_0/
\pi$ = 0.8, 0.4, 0.2, 0.02, and 0.002, in order of increasing amplitude.  For
clarity, only the first two cycles of each mode are illustrated.
\smallskip
\item{Fig.~7}Orthon solution, assuming $\beta =\gamma =-0.1$. (a) Phase of
tilt;
(b) $cos2\phi$, proportional to the `LTT strain', $e_-=e_{11}-e_{22}$; (c)
$sin2\phi$, proportional to the orthorhombic shear strain $e_{12}$.
\smallskip
\item{Fig.~8}Tetron solution, with same parameters as Fig. 7.
\smallskip
\item{Fig.~9}Family of tetron solutions, illustrating discontinuous jump in
$\alpha$ for $\gamma >\gamma_c =0.25$.  For all curves, $\hat\beta =-1$. Values
of $\gamma$ are, from right to left, 0.05, 0.1, 0.2, 0.251, 0.26, 0.3, 0.4,
0.8,
2, and 10 ($\gamma =100$ is nearly indistinguishible from $\gamma =10$.)
Filled circles show limits of jump in $\alpha$.
\smallskip
\item{Fig.~10}Local energy per unit cell of waves of Fig. 6, assuming
$\eta_{ep}
=2$, $\eta_2=0$.  For simplicity, only one half cycle of period (corresponding
to a single kink) is illustrated.
\smallskip
\item{Fig.~11}Local energy per unit cell of single (100) soliton, using
parameters appropriate to LSCO: $\eta_{ep}=8$, $\eta_2=0$, $E_0=5.5meV$.  Solid
line = total energy, Eq. 17; dotdashed line = contribution of $\hat\beta_e$;
long dashed line = contribution of $\hat\gamma_e$;
short dashed line = contribution of $\hat\delta_e$;
dotted line = contribution of $-3\phi^{\prime 2}/8$.
\smallskip
\item{Fig.~12}Integrated domain wall energy $E_w$ per column, as a function of
wall spacing, $d_w$, both for orthon (solid
lines) and tetron configurations (dashed lines), for $\eta_{ep}=8$ or 2.
\smallskip
\item{Fig.~13}Microscopic domain wall configurations, for the calculations of
Appendix II, for three different wall thicknesses.
\smallskip
\item{Fig.~14} (a) JT energy, $H_{epn}$. Solid line = Eq. A2b, $\nu =0$; dashed
line = Fourier approximation, retaining only the $n=0,1$ terms. (b) $\partial
H_{epn}/\partial\phi_n$.  Solid line = Eq. A4b, $\nu =0$; dashed line = Fourier
approximation, $n=1$ term only; dotted line = two term ($n=1,2$) Fourier
representation.
\smallskip
\end